# Holographic bounce cosmological models induced by viscous dark fluid from a generalized non-singular entropy function

*E. Elizalde[1]*

*[1]Institute of Space Science, ICE/CSIC and IEEC,
Campus UAB, C/Can Magrans, s/n, 08193
Bellaterra, Barcelona, Spain*

*A. V. Yurov[2]*

*[2]Institute of High Technology
Baltic Federal University the name Immanuel Kant (BFU)
Kaliningrad 236041, Russia*

*A. V. Timoshkin[3,4]*

*[3]Institute of Scientific Research and Development,
Tomsk State Pedagogical University (TSPU), 634041 Tomsk, Russia*

*[4]Lab. for Theor. Cosmology, International Centre of Gravity and Cosmos,
Tomsk State University of Control Systems and Radio electronics
(TUSUR), 634050 Tomsk, Russia*

elizalde@ice.csic.es
aiurov@kantiana.ru
alex.timosh@rambler.ru

Bounce cosmological models containing a dark viscous fluid in a spatially flat Friedmann-Robertson-Walker (FRW) universe are considered. The universe evolution is described in terms of generalized equation of state (EoS) parameters, in presence of the bulk viscosity. Entropic cosmology plays a key role in the discussion, and the matter bounce behavior is described based on a non-singular, generalized entropy function, recently proposed by Odintsov and Paul [1]. Three different forms for the scale factor are investigated: an exponential, a power-law, and a double-exponential function, respectively. Appropriate bounce cosmological models are formulated, via the relevant parameters of the modified EoS, and analytical expressions for the corresponding infrared cut-off are obtained, via the particle horizon. Results are displayed in holographic form, making use of generalized holographic cut-offs first introduced by Nojiri and Odintsov [49]. In addition, the viability of the corresponding bounce cosmological models is investigated, taking into account the actual thermodynamic properties of our universe, by means of a no-singular, generalized entropy function. In the asymptotic case, an expression for the generalized entropy is obtained, which remarkably has the additivity property.

## I. Introduction

The discovery of the accelerated expansion of our universe, based on astronomical observations of the teams led by A. Riess and S. Perlmutter [2, 3], led to a further development of theoretical cosmology and to the appearance of brand-new cosmological models. Cosmic acceleration can be described either by using dark energy [4] or, alternatively, by a theory of modified gravity [5]. For the accelerated expansion of the universe to occur, dark energy must have undergone a strong negative pressure phase, which can be characterized by the thermodynamic parameter in the EoS, namely $\omega = \dfrac{p}{\rho}$, where $\rho$ is the dark energy density and $p$ the dark energy pressure. According to present observational data, the value of $\omega$ is $-1.04^{+0.09}_{-0.10}$ [6].

Various possible scenarios for the late evolution of the universe, as predicted by the theory, have been extensively discussed in the literature, namely for the Big Rip [7, 8], Little Rip [9–17], Pseudo-Rip [18], and Quasi-Rip [19, 20] cosmologies.

On top of the widely considered cosmological models for the early universe, interesting additional scenarios, involving a cyclic cosmology or a bounce cosmology are also possible [21-30]. In physical terms, bounce cosmology is characterized by the fact that, at the beginning of time, the universe, which is already filled with matter, contracts in an era of accelerated collapse, and then suddenly bounces, without the appearance of any particular singularity in the process. After the bounce, a phase of accelerated, matter-dominated universe expansion begins, possibly giving rise to a cyclic universe.

In Refs. [31, 32], Nojiri, Odintsov and Faraoni introduced a generalized entropy function depending on four parameters. It made possible to explain the early inflation epoch and the late universe era of dark energy dominance in a unified way. According to entropic cosmology, Friedman's equations are a consequence of the fundamental laws of thermodynamics, since it is the entropy function, which actually generates the energy density and pressure in these equations. To be noted is that the original, generalized entropy function becomes singular during the cosmic evolution of the universe, namely when the Hubble parameter tends to zero. This can possibly occur in bounce cosmology, where the Hubble parameter "disappears" at the instant of bounce.

New entropy functions, generalizing the Tsallis [33], Barrow [34], Renyi [35], Kaniadakis [36], Sharma-Mittal [37], and Loop Quantum Gravity entropies [38] –and which turn out to be non-singular throughout the entire cosmic evolution of the universe– were recently proposed in Ref. [1]. The non-singularity of the entropy function is absolutely necessary for describing the bounce cosmology. Furthermore, it has been shown that the theoretical predictions obtained in entropic cosmology, for the spectral parameters of inflation, are in good agreement with recent astronomical observations of the Planck satellite, for all relevant entropy and energy ranges [39, 40].

In this article, we will make use of a non-singular, generalized entropy function in order to formulate the analytical conditions necessary for the appearance of bounce cosmology, under the form of an exponential, a power-law and a double-exponential function, respectively, in terms of the thermodynamic parameter and of the bulk viscosity, in the spatially flat Friedman-Robertson-Walker metric. The energy conservation law will be presented in holographic form. Moreover, we will formulate a condition for the usefulness of bounce cosmology, which duly takes into account the thermodynamic properties of our universe.

For the generalized entropy function, defined by formulas (2) and (3), the important issue of its additivity arises. Violation of the crucial additivity property in generalized entropies usually follows from the presence of quantum gravity effects, which cannot be neglected near the singularity. It is here proposed, to modify the generalized entropy formula by taking advantage of the asymptotic case corresponding to low energies and small values of the Hubble function. On the basis of Cram's theory for multi-soliton potentials, it will be shown below that the entropy function can actually recover the additivity property.

## II. Singularity-free generalized entropy function

As was shown in [31, 32], a useful, general expression for the entropy function, in terms of four positive parameters, $\alpha_+, \alpha_-, \beta, \gamma$, reads

$$S_g^{(s)}(\alpha_+, \alpha_-, \beta, \gamma) = \frac{1}{\gamma}\left[\left(1 + \frac{\alpha_+}{\beta}S\right)^{\beta} - \left(1 + \frac{\alpha_-}{\beta}S\right)^{-\beta}\right], \qquad (1)$$

where $S = \dfrac{\pi}{GH^2}$ is the Bekenstein-Hawking entropy, $H$ the Hubble function, and $G$ Newton's gravitational constant. The function $S_g^{(s)}$ (1) reduces to all well-known entropy functions (namely, those of Bekenstein-Hawking, Tsallis, Barrow, Renyi, Kaniadakis, Sharma-Mittal, and the one used in Loop Quantum Gravity), for suitable choices of the parameters.

The general entropy function Eq. (1) can be written in the equivalent form [33]

$$S_g^{(s)}(\alpha_+, \alpha_-, \beta, \gamma) = \frac{1}{\gamma}\left[\left(1 + \frac{\pi\alpha_+}{\beta GH^2}\right)^{\beta} - \left(1 + \frac{\pi\alpha_-}{\beta GH^2}\right)^{-\beta}\right]. \qquad (2)$$

The function $S_g^{(s)}$ becomes singular (diverges) at the instant of bounce, when the Hubble function of the universe tends to zero. Consequently, the generalized entropy function (1) is actually devoid of physical meaning, in bounce cosmology.

A more general entropy function, which also includes all the already mentioned ones, and which, in addition, is non-singular for the entire cosmological evolution of the universe, was proposed in [1]:

$$S_g(\alpha_+, \alpha_-, \beta, \gamma) = \frac{1}{\gamma}\left\{\left[1 + \frac{1}{\varepsilon}\tanh\left(\frac{\varepsilon\pi\alpha_+}{\beta GH^2}\right)\right]^{\beta} - \left[1 + \frac{1}{\varepsilon}\tanh\left(\frac{\varepsilon\pi\alpha_-}{\beta GH^2}\right)\right]^{-\beta}\right\}. \qquad (3)$$

As can be easily seen here, for $H \to 0$, $\tanh\left(\dfrac{\varepsilon\pi\alpha_\pm}{\beta GH^2}\right) \to 1$, and the function $S_g$ becomes finite at the moment of bounce, namely

$$S_g(\alpha_+, \alpha_-, \beta, \gamma) = \frac{1}{\gamma}\left[\left(1 + \frac{1}{\varepsilon}\right)^{\beta} - \left(1 + \frac{1}{\varepsilon}\right)^{-\beta}\right]. \qquad (4)$$

The presence of an extra parameter should be here noted. In fact, for the nonsingular entropy (3), the minimum set of parameters involved is five (see Ref. [1]).

The generalized nonsingular entropy, $S_g$, yields the corresponding energy density and pressure in the modified Friedmann equation. The energy density is given by [1]

$$\rho_g = \frac{3}{k^2}\left[H^2 - f(H; \alpha_+, \alpha_-, \beta, \gamma, \varepsilon)\right]. \qquad (5)$$

The modified Friedmann equation associated to the generalized entropy, reads

$$H^2 = \frac{k^2}{3}\left(\rho + \rho_g\right), \qquad (6)$$

where $H = \dfrac{\dot{a}(t)}{a(t)}$ is the Hubble function, $k^2 = 8\pi G$ Einstein's gravitational constant, $\rho$ the dark energy density, and $\rho_g$ the energy density corresponding to the entropy function, $S_g$. Also, $\dot{a}$ denotes derivative with respect to the cosmic time, $t$. Eq. (6) is similar to the usual Friedmann equation with total energy density $\rho_t = \rho + \rho_g$.

The general form of the function $f(H; \alpha_+, \alpha_-, \beta, \gamma, \varepsilon)$ in the approximation $GH^2 \ll 1$, the Hubble parameter being less than the Planck scale, was obtained in [1]. This approximation is valid throughout the entire evolution of the universe, and

$$f \equiv f\left(H; \alpha_+, \alpha_-, \beta, \gamma, \varepsilon\right) = \frac{4}{\gamma} H^2 \left\{ \alpha_+ \left(1 + \frac{1}{\varepsilon}\right)^{\beta-1} f_+ + \alpha_- \left(1 + \frac{1}{\varepsilon}\right)^{-\beta-1} f_- \right\}. \qquad (7)$$

Here, we have introduced the notation

$$f_\pm \equiv f_\pm\left(H; \alpha_\pm, \beta, \varepsilon\right) = \exp\left(-\frac{2\varepsilon\pi\alpha_\pm}{\beta G H^2}\right) + \left(\frac{2\varepsilon\pi\alpha_\pm}{\beta G H^2}\right) Ei^{(\pm)}, \qquad (8)$$

where $Ei^{(\pm)} = Ei\left(-\dfrac{2\varepsilon\pi\alpha_\pm}{\beta G H^2}\right)$.

The function $f$ incorporates the influence of the thermodynamic properties of the universe through the nonsingular generalized entropy on its evolution. From Eqs. (5) and (6), it follows that, in a nonsingular generalized entropy model, the dark energy density is modified, and takes the form

$$\rho = \frac{3}{k^2} f. \qquad (9)$$

Here the factor $f$ corrects the dark energy density, by taking into account the properties of the generalized entropy function.

Before considering now the bounce scenario, let us recall some basic theoretical principles.

### III. Application of the holographic principle to viscous entropic cosmology

Consider a homogeneous and isotropic, spatially flat Friedmann-Robertson-Walker (FRW) metric

$$ds^2 = -dt^2 + a^2\left(t\right) \sum_{i=1,2,3} \left(dx^i\right)^2, \qquad (10)$$

where $a\left(t\right)$ is the scale factor. We assume now that the universe consists of this one-component viscous fluid, and consider the energy conservation equation in the form

$$\dot{\rho} + 3H\left(p + \rho\right) = 0. \qquad (11)$$

The pressure $p$ of the viscous dark fluid is given in the form of a generalized (EoS) in flat FRW space time [41]

$$p = \omega\left(\rho, t\right)\rho - 3H\zeta\left(H, t\right), \qquad (12)$$

where $\omega\left(\rho, t\right)$ is a thermodynamic parameter and $\zeta\left(H, t\right)$ is the bulk viscosity, which depends on the Hubble function and on time, $t$. From thermodynamic considerations, it follows that $\zeta\left(H, t\right) > 0$.

The derivative of the function $f$

$$\dot{f} = 8\frac{\dot{H}}{H} \left\{ \frac{f}{4} - \frac{2\pi\varepsilon}{\gamma\beta G} \left[ \alpha_+^2 \left(1 + \frac{1}{\varepsilon}\right)^{\beta-1} Ei^{(+)} + \alpha_-^2 \left(1 + \frac{1}{\varepsilon}\right)^{-\beta-1} Ei^{(-)} \right] \right\}, \qquad (13)$$

is introduced into (13) to calculate the derivative of the dark energy density (9)

$$\dot{\rho} = \frac{24\dot{H}}{k^2 H} \left\{ \frac{f}{4} - \frac{2\varepsilon\pi}{\gamma\beta G} \left[ \alpha_+^2 \left(1 + \frac{1}{\varepsilon}\right)^{\beta-1} Ei^{(+)} + \alpha_-^2 \left(1 + \frac{1}{\varepsilon}\right)^{-\beta-1} Ei^{(-)} \right] \right\}. \qquad (14)$$

Next, taking into account (12) and (13), the energy conservation law for the nonsingular entropic cosmological model can be written as

$$\frac{8}{3}\left\{\frac{1}{4}f - \frac{2\varepsilon\pi}{\gamma\beta G}\left[\alpha_+^2\left(1+\frac{1}{\varepsilon}\right)^{\beta-1}Ei^{(+)} + \alpha_-^2\left(1+\frac{1}{\varepsilon}\right)^{-\beta-1}Ei^{(-)}\right]\right\}\frac{\dot{H}}{H^2} +$$
$$+(\omega+1)f - k^2 H\zeta(H,t) = 0 \qquad (15)$$

Let us now consider the special case when the entropy parameters are $\beta = 1$, $\alpha_+ = \alpha_- = \alpha$, and denote $\theta \equiv \frac{2\varepsilon\pi\alpha}{G}$. Then, in Eq. (15), the function is converted into $\tilde{f}$, namely

$$f \to \tilde{f} = \frac{4\alpha}{\gamma}\left[1+\left(\frac{\varepsilon}{1+\varepsilon}\right)^2\right]H^2\Phi(H). \qquad (16)$$

Here, we have introduced the function $\Phi(H) = \exp\left(-\frac{\theta}{H^2}\right) + \frac{\theta}{H^2}Ei\left(-\frac{\theta}{H^2}\right)$, where $\theta = \frac{2\pi\alpha\varepsilon}{G}$.

The energy-conservation equation (14) takes the form

$$\frac{8}{3}\frac{\alpha}{\gamma}\left[1+\left(\frac{\varepsilon}{1+\varepsilon}\right)^2\right]\left\{\exp\left(-\frac{\theta}{H^2}\right)\dot{H} + \frac{3}{2}(\omega+1)H^2\Phi(H)\right\} - k^2 H\zeta(H,t) = 0. \qquad (17)$$

And for the bulk viscosity, we obtain

$$\zeta(H,t) = \frac{8\alpha}{3\gamma k^2 H}\left[1+\left(\frac{\varepsilon}{1+\varepsilon}\right)^2\right]\left\{Ei\left(-\frac{\theta}{H^2}\right)\ddot{H} + \frac{3}{2}(\omega+1)H^2\Phi(H)\right\}. \qquad (18)$$

In the following we will study bounce cosmological models.

### IV. Holographic bounce cosmology corresponding to generalized nonsingular entropy

In this section, we will consider bounce cosmological models where the scale factor is described by an exponential, a power-law, or a double exponential function, respectively. Similar bounce cosmologies, which take into account dark fluid viscosity have been considered in [42, 43]. The motivation for describing the bounce in terms of a viscous fluid is that the inclusion of viscosity allows to achieve better agreement between the theoretical models and the astronomical observations [44-46].

The novelty of this study lies in the description of the behavior of the bounce cosmology by means of the thermodynamic properties of the universe based on a nonsingular generalized entropy function.

We should here recall the main statements of the holographic principle, in the terminology proposed by Li [47]. According to the holographic principle, all physical quantities within the universe, including the dark energy density, can be described by certain values at the boundary of space time [48]. In a holographic description, all physical quantities are represented through the horizon cut-off radius. According to the generalized model introduced in [49], the holographic energy density is inversely proportional to the squared infrared cut-off, $L_{IR}$

$$\rho_{hol} = 3c^2 k^2 L_{IR}^{-2}, \qquad (19)$$

where $k^2 = 8\pi G$ is Einstein's gravitational constant, being $G$ Newton's gravitational constant and $c$ a non-dimensional, positive constant. Within this description of dark energy, the horizon cut-off radius corresponds to the infrared cut-off.

Such infrared cut-off radius, describing the accelerated expansion of the universe, can be identified with the size of the particle horizon, $L_p$, or with the size of the event horizon, $L_f$. They are defined, respectively, as [3]

$$L_p(t) \equiv a(t)\int_0^t \frac{dt'}{a(t')}, \quad L_f(t) \equiv a(t)\int_t^\infty \frac{dt'}{a(t')}, \qquad (20)$$

where a($t$) is the scale factor.

In addition, the infrared boundary is given by a combination of the parameters $L_p$ and $L_f$, as well as their derivatives, and of the Hubble function, the scale factor, and corresponding derivatives; namely

$$L_{IR} = L_{IR}\left(L_p, \dot{L}_p, \ddot{L}_p, ..., L_f, \dot{L}_f, \ddot{L}_f, ..., a, ..., H, \dot{H}, \ddot{H}, ...\right). \tag{21}$$

However, not every choice of the infrared radius can lead to an accelerated expansion of the universe. As a consequence, the choice of the infrared cut-off radius is not arbitrary.

We will describe the holographic bounce in terms of the particle horizon, $L_p$. The Hubble function can be expressed in terms of the particle horizon and its derivatives, as follows [50]:

$$H = \frac{\dot{L}_p - 1}{L_p} \quad , \quad \dot{H} = \frac{\ddot{L}_p}{L_p} - \frac{\dot{L}_p^2}{L_p^2} + \frac{\dot{L}_p}{L_p^2} \,. \tag{22}$$

Then, by using (22), the energy conservation law (17) in the holographic language takes the form

$$\frac{8}{3}\frac{\alpha}{\gamma}\left[1+\left(\frac{\varepsilon}{1+\varepsilon}\right)^2\right]\left\{\begin{array}{l}\exp\left(-\dfrac{\theta}{\left(\dfrac{\dot{L}_p-1}{L_p}\right)^2}\right)\left(\dfrac{\ddot{L}_p}{L_p}-\dfrac{\dot{L}_p^2}{L_p^2}+\dfrac{\dot{L}_p}{L_p^2}\right)+\\[20pt]+\dfrac{3}{2}(\omega+1)\left(\dfrac{\dot{L}_p-1}{L_p}\right)^2\Phi\left(\dfrac{\dot{L}_p-1}{L_p}\right)\end{array}\right\}-k^2\left(\dfrac{\dot{L}_p-1}{L_p}\right)\zeta\left[\left(\dfrac{\dot{L}_p-1}{L_p}\right),t\right]=0\,. \tag{23}$$

In this way, we have shown the equivalence of the dark fluid description with the specific holographic cut-off introduced in [49].

Let us now rewrite the bulk viscosity (18) in the holographic form

$$\zeta\left(H,t\right)=\frac{8\alpha}{3\gamma k^2 \dfrac{\dot{L}_p-1}{L_p}}\left[1+\left(\frac{\varepsilon}{1+\varepsilon}\right)^2\right].$$

$$\cdot\left\{\exp\left(-\dfrac{\theta}{\left(\dfrac{\dot{L}_p-1}{L_p}\right)^2}\right)\left(\dfrac{\ddot{L}_p}{L_p}-\dfrac{\dot{L}_p^2}{L_p^2}+\dfrac{\dot{L}_p}{L_p^2}\right)+\dfrac{3}{2}(\omega+1)\left(\dfrac{\dot{L}_p-1}{L_p}\right)^2\Phi\left(\dfrac{\dot{L}_p-1}{L_p}\right)\right\}. \tag{24}$$

In what follows, we will discuss three different cosmological models with bounce via the bulk viscosity.

### a. Exponential model

We first consider a bounce cosmological model with scale factor, $a$, given by the exponential function [24]

$$a(t) = a_0 \exp\left[\delta(t-t_0)^2\right], \qquad (25)$$

where $a_0$ and $\delta$ are positive (dimensional) constants and $t_0$ is a fixed bounce time. The Hubble function behaves as

$$H(t) = 2\delta(t-t_0). \qquad (26)$$

At time $t$, previous to the bounce time $t_0$ $(t < t_0)$, the scale factor decreases, and the universe contracts. At the moment of the bounce, $t_0$, the scale factor is constant, $a_0 = a(t_0)$. After the bounce time $t_0$ $(t > t_0)$, the scale factor increases and the universe begins to expand. Now, using (18), we can obtain the particle horizon $L_p$, with the result

$$L_p = a \int_0^t \frac{dt}{a} = \frac{1}{2}\sqrt{\frac{\pi}{\delta}} \exp\left[\delta(t-t_0)^2\right] erf\left[\delta(t-t_0)^2\right], \qquad (27)$$

being $erf\left[\delta(t-t_0)^2\right]$ the usual probability integral.

If we introduce the particle horizon $L_p$, given in Eq. (26), into Eq. (23), we obtain the bulk viscosity in holographic form, corresponding now to the bounce exponential model, which contains the additional contribution coming from the generalized non-singular entropy.

Consider, further, the simple case, in which the thermodynamic parameter in the EoS is constant, $\omega(\rho,t) = \omega_0$, and the bulk viscosity is directly proportional to the Hubble function, $H$: $\zeta(H,t) = 3\tau H$, with $\tau$ a positive constant. For the EoS (12), we obtain

$$p = \frac{3}{k^2}\left(\omega_0 \tilde{f} - 3\tau k^2\right)H^2, \qquad (28)$$

where the function $\tilde{f}$ (15) corrects the influence of the nonsingular generalized entropy. Then, the energy conservation law (22) in the present model can be rewritten as

$$\frac{8}{3}\frac{\alpha}{\gamma}\left[1+\left(\frac{\varepsilon}{1+\varepsilon}\right)^2\right]\left\{ \begin{array}{l} \exp\left(-\dfrac{\theta}{\left(\dfrac{\dot{L}_p-1}{L_p}\right)^2}\right)\left(\dfrac{\ddot{L}_p}{L_p}-\dfrac{\dot{L}_p^2}{L_p^2}+\dfrac{\dot{L}_p}{L_p^2}\right)+ \\ +\dfrac{3}{2}(\omega_0+1)\left(\dfrac{\dot{L}_p-1}{L_p}\right)^2\Phi\left(\dfrac{\dot{L}_p-1}{L_p}\right) \end{array} \right\} -3\tau k^2\left(\dfrac{\dot{L}_p-1}{L_p}\right)^2 = 0. \quad (29)$$

As a result, we have obtained a quite simple viscous bounce model, based upon the assumption (25) for the scale factor.

### b. Power-law model

As a second example, we consider the bounce cosmological model corresponding to a scale factor that is expressed by a power function, namely [24]

$$a(t) = a_0 + \lambda(t-t_0)^2, \qquad (30)$$

where $a_0, \lambda$ are constant, and $t_0$ is a fixed bounce time. Here, as before, $t_0$ is the point of change of the evolution mode, the bounce.

The Hubble function has the form

$$H(t) = \frac{2\lambda(t - t_0)}{a_0 + \lambda(t - t_0)^2},\tag{31}$$

And for the particle horizon, $L_p$, one obtains

$$L_p = \frac{1}{\sqrt{a_0\lambda}}\left[a_0 + \lambda(t - t_0)^2\right]\arctan\sqrt{\frac{\lambda}{a_0}}(t - t_0).\tag{32}$$

In this case, using (29) and (23), we can represent the bounce power-law model in holographic form, via the bulk viscosity taking into account the thermodynamic properties of the universe.

Consider now the case of constant bulk viscosity, $\zeta(H,t) = \zeta_0 > 0$. Also, let us assume that the thermodynamic parameter, $\omega$, is a linear function of the energy density, as

$$\omega(\rho,t) = A_0\rho - 1 = \frac{3A_0}{k^2}\tilde{f} - 1 = \frac{12A_0\alpha}{\gamma k^2}\left[1 + \left(\frac{\varepsilon}{1+\varepsilon}\right)^2\right]H^2\Phi(H) - 1,\tag{33}$$

where $A_0$ is a positive constant having dimensions of $cm^4$ in geometric units. The energy conservation law in the holographic language takes the form

$$\frac{8}{3}\frac{\alpha}{\gamma}\left[1 + \left(\frac{\varepsilon}{1+\varepsilon}\right)^2\right]\left\{\exp\left(-\frac{\theta}{\left(\frac{\dot{L}_p - 1}{L_p}\right)^2}\right)\left(\frac{\ddot{L}_p}{L_p} - \frac{\dot{L}_p^2}{L_p^2} + \frac{\dot{L}_p}{L_p^2}\right) + \right.$$
$$\left. + \frac{18A_0\alpha}{\gamma k^2}\left[1 + \left(\frac{\varepsilon}{1+\varepsilon}\right)^2\right]\left[\left(\frac{\dot{L}_p - 1}{L_p}\right)^2\Phi\left(\frac{\dot{L}_p - 1}{L_p}\right)\right]^2\right\} - \zeta_0 k^2\left(\frac{\dot{L}_p - 1}{L_p}\right) = 0.\tag{34}$$

This is here the power-law realization of the holographic principle, which describes the early universe for the viscous fluid.

### c. Double exponential model

Next, we explore the case of a sum of different exponential functions, namely a scale factor of the form [24]

$$a(t) = \exp(Y) + \exp(Y^2),\tag{35}$$

where $Y = \frac{t}{\bar{t}}$ and $\bar{t}$ is the reference time. From this expression, we have

$$H(t) = \frac{1 + 2Y\exp(Y^2 - Y)}{\bar{t}\left[1 + \exp(Y^2 - Y)\right]}.\tag{36}$$

Near the bounce time, when $t \to 0$, a calculation of the particle horizon yields

$$L_p = a \int_0^t \frac{dt}{e^{\left(\frac{t}{\bar{t}}\right)^2} + e^{\left(\frac{t}{\bar{t}}\right)^4}} = a \int_0^t \frac{dt}{2 + \left(\frac{t}{\bar{t}}\right)^2 + \left(\frac{t}{\bar{t}}\right)^4} =$$

$$= \frac{a\bar{t}}{3\sin\alpha} \left( \sin\frac{\alpha}{2} \ln \frac{Y + \sqrt[4]{8}\cos\frac{\alpha}{2}\sqrt{Y} + \frac{\sqrt{2}}{2}}{Y - \sqrt[4]{8}\cos\frac{\alpha}{2}\sqrt{Y} + \frac{\sqrt{2}}{2}} + 2\cos\frac{\alpha}{2} \arctan \frac{Y - \frac{\sqrt{2}}{2}}{\sqrt{2}\sin\frac{\alpha}{2}\sqrt{Y}} \right), \quad (37)$$

where $\cos\alpha = -\dfrac{1}{2\sqrt{2}}$. And substituting Eq. (37) into Eq. (24), we obtain the holographic description of a bounce double exponential model via bulk viscosity.

To conclude, let us now represent the energy conservation law in this model, in holographic form, taking the thermodynamic parameter and the bulk viscosity in the EoS to be constant, say: $\omega(\rho, t) = \omega_0$, $\zeta(H, t) = \zeta_0$. We then obtain

$$\frac{8}{3}\frac{\alpha}{\gamma}\left[1 + \left(\frac{\varepsilon}{1+\varepsilon}\right)^2\right] \left\{ \begin{array}{l} \exp\left(-\dfrac{\theta}{\left(\dfrac{\dot{L}_p - 1}{L_p}\right)^2}\right) \left(\dfrac{\ddot{L}_p}{L_p} - \dfrac{\dot{L}_p^2}{L_p^2} + \dfrac{\dot{L}_p}{L_p^2}\right) + \\ + \dfrac{3}{2}(\omega_0 + 1)\left(\dfrac{\dot{L}_p - 1}{L_p}\right)^2 \Phi\left(\dfrac{\dot{L}_p - 1}{L_p}\right) \end{array} \right\} - \zeta_0 k^2 \left(\dfrac{\dot{L}_p - 1}{L_p}\right) = 0. \quad (38)$$

This equation is the result of the double exponential model, as following from a viscous holographic model.

## V. Influence of a generalized entropy function, on the viability of bounce cosmology

In previous sections the singular-free generalized entropy function (3) and its application to describe bounce cosmological models were considered. The question is now: how does the generalized entropy function affect the viability of the bounce cosmological models that are consistent with the constraints of the astronomical observations?

To answer it, we apply the first law of thermodynamics to the non-singular generalized entropy function $S_g = S_g(S)$, which is a function of the Bekenstein-Hawking entropy, $S = \dfrac{\pi}{GH^2}$. We obtain,

$$T dS_g = dQ, \quad (39)$$

which, in an equivalent form, can be written as

$$T\left(\frac{\partial S_g}{\partial S}\right) dS = dQ. \quad (40)$$

The energy flux $E$ is equal to the heat $Q$ on the cosmological horizon, which has radius $r_H = \dfrac{1}{H}$. That is,

$$dQ = -dE = -\frac{4}{3}\pi r_h^3 \dot{\rho} dt = -\frac{4\pi}{3H^3}\dot{\rho} dt = \frac{4\pi}{H^2}(p + \rho) dt. \quad (41)$$

Then, using the Hawking temperature, $T = \dfrac{1}{2\pi r_h} = \dfrac{H}{2\pi}$ [51], and the expression for the Bekenstein-Hawking entropy, we can rewrite the thermodynamic equation (40) under the form

$$\dot{H}\left(\frac{\partial S_g}{\partial S}\right) = -4\pi G\left(p + \rho\right).$$ (42)

For further analysis, it is assumed that the matter field and the cosmological constant are absent, i.e. $\rho = p = \Lambda = 0$. We now take (as before) the parameters $\alpha_+ = \alpha_- = \alpha$ and $\beta = 1$. Then, without loss of generality, after calculating the derivative of the function $S_g$, Eq. (42) reads

$$\frac{\alpha}{\gamma}\sec h^2\left(\frac{\pi\alpha\varepsilon}{GH^2}\right)\left\{1 + \left[1 + \frac{1}{\varepsilon}\tanh\left(\frac{\pi\alpha\varepsilon}{GH^2}\right)\right]^{-2}\right\}\dot{H} = 0.$$ (43)

The solution of Eq. (43) is the Hubble parameter $H = 0$, which does not describe the correct evolution of the universe. Therefore, we are bound to consider a more general entropy function, $S_g$, with time-varying parameters.

In the following, let us find the conditions under which by changing the parameters of the generalized entropy function we are led to a correct description of the universe evolution. To do this, we will see that it is enough to just vary one of the parameters over time, $t$, for example $\gamma$, while the other four parameters $\alpha_+, \alpha_-, \beta, \varepsilon$ may remain constant. Quite reasonably, the change of a parameter over time may be attributed to a quantum gravity effect.

Thus, we set

$$\gamma = \gamma\left(N\right),$$ (44)

where $N$ is the number of e-folds of the universe, which stays here for time. The modified Friedmann equation corresponding to the generalized entropy function, $S_g$, reads

$$\frac{2\pi}{G}\left(\frac{\partial S_g}{\partial S}\right)\frac{H'(N)}{H^3} = \left(\frac{\partial S_g}{\partial \gamma}\right)\gamma'(N),$$ (45)

where we take $\rho = p = 0$, to describe the bounce universe, and the prime denotes $d / dN$. Substituting in Eq. (45) the expression for the generalized entropy, $S_g$ (3), we get

$$-\frac{2\pi\alpha}{G}\frac{H'(N)}{H^3}\sec h^2\left(\frac{\pi\alpha\varepsilon}{GH^2}\right)\frac{\left[1 + \dfrac{1}{\varepsilon}\tanh\left(\dfrac{\pi\alpha\varepsilon}{GH^2}\right)\right]^2 + 1}{\left[1 + \dfrac{1}{\varepsilon}\tanh\left(\dfrac{\pi\alpha\varepsilon}{GH^2}\right)\right]^2 - 1} = \frac{\gamma'(N)}{\gamma(N)}.$$ (46)

From here it follows that, since $\gamma$ changes over time, then t$\gamma'(N) \neq 0$ and the Hubble function is not a constant. Therefore, a non-singular bounce model based on a generalized entropy function does describe a realistic universe.

Using the definition of the Bekenstein-Hawking entropy, $S = \dfrac{\pi}{GH^2}$, Eq. (46) takes the form

$$\alpha\sec h^2\left(\alpha\varepsilon S\right)\frac{\left[1 + \dfrac{1}{\varepsilon}\tanh\left(\alpha\varepsilon S\right)\right]^2 + 1}{\left[1 + \dfrac{1}{\varepsilon}\tanh\left(\alpha\varepsilon S\right)\right]^2 - 1}dS = \frac{\gamma'(N)}{\gamma(N)}dN.$$ (47)

After integration, one gets [1]

$$1+\frac{1}{\varepsilon}\tanh(\alpha\varepsilon S)-\left[1+\frac{1}{\varepsilon}\tanh(\alpha\varepsilon S)\right]^{-1}=\gamma(N).\qquad(48)$$

Eq. (46) defines the Hubble function in terms of the e-fold number, $H=H(N)$, in an implicit form. Rewriting Eq. (48) in the form

$$\tanh\left(\frac{\pi\alpha\varepsilon}{GH^2}\right)=\frac{\gamma(N)+\sqrt{\gamma^2(N)+4}}{2}-1,\qquad(49)$$

we see that the time evolution of the Hubble function $H(N)$ depends on the form of the variable parameter $\gamma(N)$.

Due to the fact that the Hubble function appears in quadratic power in Eq. (49), its solution contains both positive and negative branches of $H(N)$. This allows for the possibility of a symmetric bounce in the models here discussed of a nonsingular entropy cosmology.

Now, we will check the fulfillment of condition (49) for the cosmological models with bounce discussed in section IV. With this, we will manage to establish their possible viability.

## VI. Viability of bounce cosmological models in entropy cosmology

In this section we will investigate the validity of the bounce models considered above, namely the exponential, power and double exponential ones, in the presence of viscosity, taking advantage of the generalized non-singular entropy function, $S_g$.

### a. Feasibility of the exponential bounce

In this bounce model, the scale factor and the Hubble function have the form given in Eqs. (25) and (26), respectively. Using the fact that $N=\ln a$, we can obtain a relation between the cosmic time and the e-fold number, $N$, namely

$$t(N)-t_0=\pm\sqrt{\frac{N}{\delta}},\qquad(50)$$

where we have set $a_0=1$. Then, the Hubble function in terms of the e-fold number reads

$$H(N)=\pm2\sqrt{\delta}N^{\frac{1}{2}}.\qquad(51)$$

It follows from Eqs. (50) and (51) that the cosmic time and the Hubble function have identical signs. Thus, if the cosmic time is negative, then the Hubble function is also negative, and conversely. From a physical point of view, this corresponds to a contracting or an expanding stage for the universe evolution, the bounce time, $t_0$, corresponding to the e-fold number $N=0$. Consequently, the evolution of the universe begins from the distant past, goes through a bounce, and, subsequently, a change in the evolutionary mode occurs. In both cases, the e-fold number diverges, $N\to+\infty$.

Next, we reformulate the condition for the feasibility of a bounce model for the non-singular entropy function Eq. (48). To this end, we write the expression for $\gamma(N)$, taking into account $H(N)$

$$\gamma(N)=1+\frac{1}{\varepsilon}\tanh\left(\frac{\pi\alpha\varepsilon}{4\delta GN}\right)-\left[1+\frac{1}{\varepsilon}\tanh\left(\frac{\pi\alpha\varepsilon}{4\delta GN}\right)\right]^{-1}.\qquad(52)$$

This representation of the entropic parameter, $\gamma(N)$, proves the acceptability of the exponential bounce model, established from the singularity-free generalized entropy (3).

*b. Feasibility of a power-law bounce*

In this scenario, the scale factor and Hubble function are given by the expressions (30) and (31). The cosmic time in terms of e-fold number reads

$$t(N) - t_0 = \pm \sqrt{\frac{1}{\lambda}} \left( e^N - a_0 \right), \tag{53}$$

and the Hubble function in terms of e-fold number is

$$H(N) = \pm \sqrt{2\lambda} \sqrt{e^N - a_0} \, e^{-N}. \tag{54}$$

As in the previous case, both the cosmic time and the Hubble function have the same signs. Here the bounce time, $t_0$, corresponds to the e-fold number $N = \ln a_0$. This bounce model is also symmetric with respect to the bounce time.

Taking $\lambda = \dfrac{\pi \alpha \varepsilon}{2G}$ and $a_0 = 0$, and using the expression for $H(N)$, we can write the expression for the entropic parameter $\gamma(N)$, as

$$\gamma(N) = 1 + \frac{1}{\varepsilon} \tanh \left( e^N \right) - \left[ 1 + \frac{1}{\varepsilon} \tanh \left( e^N \right) \right]^{-1}. \tag{55}$$

As a consequence, the description of the cosmological evolution in the bounce power-law model, in the presence non-singular generalized entropy has been proven to be realistic.

*c. Feasibility of the double-exponential bounce*

In this section we explore the validity of the double-exponential bounce model in the framework of entropy cosmology. In this case, the scale factor and the Hubble function are given by Eqs. (33) and (34). By analogy with the other bounce models discussed above, we express again the cosmic time and the Hubble function through the e-fold number, $N$. This bounce model is also symmetric with respect to the bounce time.

We limit ourselves to considering the cosmological model near the moment of bounce, so that the bounce time $t \to 0$ $(Y \to 0)$. In this approximation, the cosmic time reads

$$t(N) = \frac{1}{2} \bar{t} \sqrt{4e^N - 7} \tag{56}$$

and the corresponding Hubble function

$$H(N) = \frac{1 + \sqrt{4e^N - 7} \exp \left( 4e^N - 7 + \sqrt{4e^N - 7} \right)}{\bar{t} \left[ 1 + \exp \left( 4e^N - 7 + \sqrt{4e^N - 7} \right) \right]}. \tag{57}$$

From Eq. (57), we obtain expression (50) above for the variable parameter $\gamma(N)$.

Therefore, the double-exponential bounce model corresponding to the generalized non-singular entropy function $S_g$, yields a variable Hubble function, as a consequence of the fact that the evolution of $H(N)$ does depend on the form of the entropic parameter, $\gamma(N)$. As a result, also in this case, we obtain the correct description of the universe evolution.

Next, we will investigate the possibility of restoring the additivity property of the generalized entropy.

## VII. On the question of entropy's additivity

If we interpret entropy as hidden information (number of indistinguishable microstates), then the question of the additivity of this quantity will arise. This applies both to the general case of the Bekenstein-Hawking entropy and also to the generalized entropy defined by formulas (2), (3). Indeed, for a Schwarzschild Bekenstein black hole, the entropy is proportional to the square of the mass. When two black holes merge, the final total mass equals the sum of the larger mass plus nine tenths of the smaller one (the remaining ten percent being carried away by gravitational waves). At the same time, entropy increases, but of course we cannot talk about additivity. This is not desirable, since additivity is a most important property entropy displays in thermodynamics. Precisely as a consequence of additivity, entropy is a homogeneous first-order function of all independent variables. Generally speaking, it is the enjtropy's additivity that allows to extend this concept to thermodynamic systems of any complexity. The fact that the Bekenstein-Hawking entropy and the generalized entropies (2), (3) do not have this property casts serious doubts on whether one actually has the right to use the term "entropy" in these cases.

But one can look at this situation from a different side. For instance, the fact that it is not the volume, but the area of the volume's boundary what is used to estimate the upper limit of entropy, may be definitely considered as a manifestation of the effects of quantum gravity. Indeed, the upper limit is equal to the number of Planck areas in which this area can be tiled. In other words, the classical additivity of the thermodynamic entropy should be recovered under the condition that the effects of quantum gravity are neglected. This means that we can use the fulfillment or violation of the property of entropy additivity as a criterion for the absence or presence of quantum gravity effects. Concerning cosmology, quantum effects are significant in the vicinity of a cosmological singularity, i.e. when the Hubble parameter $H \to \infty$, and they disappear when $H \to 0$. This means that for large values of the Hubble parameter (with the Hawking-Bekenstein entropy tending to zero), the generalized entropy will not have the additivity property, but with a decrease in the Hubble parameter (with a corresponding increase in the Hawking-Bekenstein entropy), including the neighborhood of the bounce, the additivity property for the generalized entropy should necessarily be restored. To demand this behavior looks quite natural.

The question then arises: is it possible to modify formula (3) so that this property is fulfilled and, at the same time, all (or most) of the advantages of expression (3) are preserved? As we will now see, this is indeed possible.

Let us set $\alpha_- = 0$, $\beta = 1$ and replace in (3) the expressions with hyperbolic tangent according to the scheme $\tanh(x) \to \left(\tanh \sqrt{x}\right)^2$. In this case, in the limit $x \to 0$ we still have the same asymptotic behavior. As a result, (3) takes the form

$$S_g(\alpha, \gamma) = \frac{1}{\gamma}\left(1 - \frac{1}{\cosh\left(\sqrt{\alpha_+ S}\right)}\right) \tag{58}$$

with $S = \dfrac{\pi}{GH^2}$.

The relation (58) suggests the use of Cram's famous formulas for further generalization. These formulas, among other things, describe multi-soliton solutions of KdV type equations. Such solutions have the following asymptotic behavior: for $t = 0$, a certain complex profile of interacting waves appear, while for $t \to +\infty$ the expression breaks down into a set of individual non-interacting solitons. For $t \to -\infty$, the picture looks similar, and the only difference between the two asymptotics $t \to \pm\infty$ is the presence of some phase shifts, which are calculated using a special procedure. But this is exactly the behavior that we may expect from entropy, which has the additivity property. Over time, the solitons diverge further and further (they all move in one direction and the higher ones have higher velocities), which means that their "interaction" disappears exponentially and, after some time, the solution is already (to arbitrarily high accuracy) simply the sum of individual solitons. If we identify each soliton with the entropy of a specific region, then the total entropy automatically becomes a simple sum of these entropies. In other words. Cram's formulas give a concrete mathematical embodiment to the property of additivity and fit the scheme and interpretation proposed above.

But, why use Cram's formulas here? There are probably many ways to come up with additive formulas. What is good about this specific choice is that formula (58) has already the form of a one-soliton solution (with the correct interpretation of the argument of the hyperbolic cosine). Our proposal is simply to move from single-soliton expressions (58) to multi-solitonic ones. This is a quite natural generalization of (58).

For the convenience of readers who may not be familiar with the theory of solitons, we will briefly explain the logic of the origin of the Cram formulas. Let $u = u(x,t)$ and the set of functions $\psi_i = \psi_i(x,t,\lambda_i)$ satisfy the equations

$$u_t - 6uu_x + u_{xxx} = 0 \qquad (59)$$

and

$$\begin{aligned} \psi_{xx} &= (u - \lambda)\psi \\ \psi_t &= 2(u + 2\lambda)\psi_x - u_x\psi \end{aligned} \qquad (60)$$

Eq. (59) is the KdV equation, whereas (60) are two linear equations whose compatibility condition is eq. (59) (so called LA pair). The next step uses the Darboux transform. Let $\psi_1$ a particular solution (60) for $\lambda = \lambda_1$. Then the LA-pair (58) is covariant under the transformation

$$\begin{aligned} \psi &\rightarrow \psi^{(1)} = \frac{\psi_x\psi_1 - \psi\psi_{1x}}{\psi_1}, \\ u &\rightarrow u^{(1)} = u - 2(\ln\psi_1)_{xx} \end{aligned} \qquad (61)$$

which can be verified by simple calculation.

If we apply (61) N times, then the Cram formulas arise, one of which is

$$u \rightarrow u^{(N)} = u - 2(\ln W_N)_{xx}, \qquad (62)$$

where $W_N$ is Cram's determinant

$$W_N = \det \begin{pmatrix} \partial_x^{N-1}\psi_N & \partial_x^{N-1}\psi_{N-1} & \cdots & \partial_x^{N-1}\psi_1 \\ \partial_x^{N-2}\psi_N & \partial_x^{N-2}\psi_{N-1} & \cdots & \partial_x^{N-2}\psi_1 \\ \vdots & & & \\ \partial_x\psi_N & \partial_x\psi_{N-1} & \cdots & \partial_x\psi_1 \\ \psi_N & \psi_{N-1} & \cdots & \psi_1 \end{pmatrix}. \qquad (63)$$

Let us briefly explain the appearance of expression (63). After N-fold iteration, the function $\psi$ takes the form

$$\psi^{(N)} = \partial_x^N\psi + \sum_{i=1}^N C_i\partial_x^{i-1}\psi, \qquad (64)$$

with some coefficients $C = C_i(x,t,\lambda_1,...,\lambda_N)$. Substituting (64) into the expression for the transformed function $u$, we find that it depends on the leading coefficient in (64). The coefficients themselves are found as follows: it is obvious that by substituting functions

$\psi_i = \psi_i(x, t, \lambda_i)$, with $i = 1, 2, \ldots, N$, into (64) we gets zero. This means that a system of N inhomogeneous linear algebraic equations arises for N unknowns $C_i$, namely

$$\sum_{i=1}^{N} C_i \partial_x^{i-1} \psi_k = -\partial_x^N \psi_k, \quad k = 1, \ldots, N \tag{65}$$

which we now solve using Cramer's method. As a result, we obtain Cramer's formula (62).

Multi-soliton solutions to equation (59) can be obtained on the vacuum background u=0. For example, a one-soliton solution is calculated using formula (62) for N=1, u=0 and $\psi_1 = \cosh(k_1 x - 4k_1^3 t)$, and $\lambda_1 = -k_1^2$. Thus

$$u^{(1)} = -\frac{k_1^2}{\cosh^2\left(k_1 x - 4k_1^3 t\right)} , \tag{66}$$

which, up to additive and multiplicative constants, coincides with (58), with the notation

$$\sqrt{\alpha_+} = k_1, \quad S = \left(x - 4k_1^2 t\right)^2 .$$

Another way to clarify this notation is: $x \equiv \sqrt{S}$ and we are using additive time shift

$$\sqrt{S} \to \sqrt{S} - 4\alpha_+ t . \tag{67}$$

To generate a two-soliton potential, two solutions of the system (60) with u=0 and $\lambda_1 = -k_1^2$, $\lambda_2 = -k_2^2 < \lambda_1$ should be used:

$$\psi_1 = \cosh\left(k_1 x - 4k_1^3 t\right), \quad \psi_2 = \sinh\left(k_2 x - 4k_2^3 t\right). \tag{68}$$

In the case of N-solitons with general positions, we have

$$\psi_{2n-1} = \cosh\left[k_{2n-1}\left(x - 4k_{2n-1}^2 t\right)\right], \quad \psi_{2n} = \sinh\left[k_{2n}\left(x - 4k_{2n}^2 t\right)\right] . \tag{69}$$

This choice guarantees a nonsingular behavior of the multi-soliton solution, provided the eigenvalues are ordered in descending order.

It remains to be noted a last circumstance: for the N-soliton expression we have N+1 free parameters, $\gamma, k_1, k_2, \ldots, k_N$. One can add N additional parameters, lying in the range from zero to one, as follows. Let denote $f_i = k_i\left(\sqrt{S} - 4k_i^2 t\right)$, then N positive parameters lying in the specified interval $\left(0 \leq \mu_i \leq 1\right)$ are introduced by replacing

$$\begin{aligned} \cosh f_i &\to \frac{1}{2}\left(e^{f_i} + \mu_i e^{-f_i}\right) \\ \sinh f_i &\to \frac{1}{2}\left(e^{f_i} - \mu_i e^{-f_i}\right) \end{aligned} . \tag{70}$$

Replacing expressions in Eqs. (69) with (70), and substituting them in (63) and (62), we finalyy obtain an expression for the generalized entropy. We see that the entropy is actually localized at N points (along the "S" axis) and is equal to the simple algebraic sum of these quantities (additivity property) for a sufficiently large value of $t$, and this "additive" entropy (up to rapidly disappearing quantum contributions) depends on $\boldsymbol{2N+1}$ free parameters.

For example, when N=2, the expression for the generalized entropy depends on five parameters and has the form

$$S_g\left(k_1, k_2, \mu_1, \mu_2, \gamma\right) = \frac{1}{\gamma}\left[1 - \frac{8\left(k_2^2 - k_1^2\right)A\left(k_1, k_2, \mu_1, \mu_2\right)}{B\left(k_1, k_2, \mu_1, \mu_2\right)}\right] \tag{71}$$

where

$$A\left(k_1, k_2, \mu_1, \mu_2\right) = \mu_1 k_1^2 \left(e^{2f_2} + \mu_2^2 e^{-2f_2}\right) + \mu_2 k_2^2 \left(e^{2f_1} + \mu_1^2 e^{-2f_1}\right) + 2\mu_1 \mu_2 \left(k_2^2 - k_1^2\right)$$

$$B\left(k_1, k_2, \mu_1, \mu_2\right) = \left[\left(k_2 - k_1\right)\left(e^{f_1+f_2} + \mu_1 \mu_2 e^{-f_1-f_2}\right) + \left(k_2 + k_1\right)\left(\mu_1 e^{f_2-f_1} + \mu_2 e^{f_1-f_2}\right)\right]^2 \quad (72)$$

It is obvious that the condition $k_2 > k_1$, $\mu_1 > 0$, $\mu_2 > 0$ guarantees the no singularity of expression (71), since $B\left(k_1, k_2, \mu_1, \mu_2\right) \neq 0$, $\forall x, t$, $x = \dfrac{1}{H}\sqrt{\dfrac{\pi}{G}}$.

Note that the parameter $t$, included as an argument, should not necessarily be identified with the time variable. It is perfectly acceptable to choose it, as a quantity canonically conjugate to the variable "time" - inverse energy. Then, large values of "t" will correspond to small values of energy and vice versa. Accordingly, at high energies both the variable $t$ and the variable $x$ (the inverse Hubble parameter) tend to zero and the generalized entropy does not appear as the sum of individual contributions. But, as already noted, this is quite natural, since higher energies mean the manifestation of quantum gravity effects; and given the low level of our understanding of these effects, it may well turn out that the concept of additive entropy simply does not work in this area. Accordingly, at low energies and small values of the Hubble parameter (i.e., large values of the $x$ variable), the dynamics become classical and the entropy must display additivity. This is perfectly described by the Cram's formulas for "multi-soliton" potentials.

## VIII. Conclusion

In this paper, we have applied the holographic principle to the study of bounce cosmology, on the base of a non-singular, generalized entropy function. We have studied three different examples, corresponding to an exponential, a power-law, and a double-exponential function, respectively. In all three cases, the dissipative dark fluid model of the universe in a homogeneous and isotropic FRW metric has been considered.

In order to investigate the evolution of the universe, we have taken advantage of the holographic principle, on the basis of a generalized model of holographic dark energy, as proposed in Ref. [49]. Within the framework of this model, using the generalized (EoS) for the dark viscous fluid, we studied the influence of the generalized non-singular entropy on the formation of bounce cosmology, represented via the bulk viscosity. The holographic energy density represents here the energy of the infrared radiation. The infrared radius $L_{IR}$ was chosen, based on the modified Nojiri-Odintsov holographic model, which is a special case of holographic dark energy. For each bounce cosmological model, an analytic expression has been obtained for the infrared radius, in terms of the particle horizon, $L_p$. The energy conservation equation has been expressed in holographic language.

It has been then shown that, as a consequence of modifying the Friedman equation by adding the energy density --which corresponds to the generalized entropy function-- corrections in the energy-conservation equation appear, which make the description of the evolution of the universe become more complete. Corrections of the thermodynamic parameters are associated with the generalized non-singular entropy function, and show up as a physical consequence in this formalism. Actually, the description of the Little Rip and Pseudo Rip cosmological models, in terms of the parameters of the EoS based on a new generalized entropy function, proposed in Ref. [32], was previously considered in Ref. [52].

We have discussed theoretical applications of entropy cosmology, based on a non-singular generalized entropy function. A necessary condition for a correct description of the evolution of the universe --in the bounce model with a non-singular generalized function $S_g$ -- is to provide the change of its entropic parameters over time. Without loss of generality, it has been here assumed that the entropic parameter $\gamma$ changes over time, while the other parameters remain constant. To this end, we have calculated the form of the entropic parameter $\gamma(N)$ in three different symmetric bounce models: an exponential, a power-law and a double-exponential bounce. Taking advantage of the fact that the change in the Hubble function, $H(N)$, is associated with the behavior of the parameter $\gamma(N)$, we have managed to prove the viability of the three bounce models here considered.

The last issue discussed in the paper is our understanding that violation of the additivity property of the generalized entropy function (3) is just a consequence of the effects of quantum gravity, which arise near the cosmological singularity. This happens for large values of the Hubble function, or near the bounce point, for small values of the Hubble function. The novel result has been proven that, as a consequence of simply modifying Eq. (3), by using basic soliton theory and the Cram formulas, it becomes possible to construct, in the asymptotic case, a mathematical expression for the generalized entropy (71), which duly displays the additivity property.

**Acknowledgements.** This work has been partially supported by MICINN (Spain), projects PID2019-104397GB-I00 and 2024AEP171, of the Spanish State Research Agency program AEI/10.13039/501100011033, by the Catalan Government, AGAUR project 2021-SGR-00171, and by the program Unidad de Excelencia María de Maeztu CEX2020-001058-M. It has been also supported by the Ministry of Science and Higher Education of the Russian Federation (agreement no. 075-02-2024-1430).